\documentclass[twocolumn]{revtex4}
\usepackage{amssymb}
\usepackage{graphicx}

\usepackage{color}

\begin{document}
\title{Random geometric graphs with general connection functions}
\date{\today}
\author{Carl P. Dettmann}
\affiliation{School of Mathematics, University of Bristol, University Walk, Bristol BS8 1TW, UK}
\author{Orestis Georgiou}
\affiliation{Toshiba Telecommuncations Research Laboratory, 32 Queens Square, Bristol BS1 4ND, UK}

\begin{abstract}
In the original (1961)  Gilbert model of random geometric graphs, nodes are placed according to a Poisson point process, and links formed between those within a fixed
range.  Motivated by wireless ad-hoc networks ``soft'' or ``probabilistic'' connection models have recently been introduced, involving a ``connection function'' $H(r)$
that gives the probability that two nodes at distance $r$ are linked (directly connect).   In many applications (not only wireless networks), it is desirable that the graph is 
connected, that is every node is linked 
to every other node in a multihop fashion. Here, the
connection probability of a dense network in a convex domain in two or three
dimensions is expressed in terms of contributions from boundary components, for a very general class of connection functions.  It turns out that only a few quantities such as
moments of the connection function appear. Good agreement is found with special cases from previous studies and with numerical simulations.
\end{abstract}

\maketitle

\section{Introduction}
\subsection{Background}
A random geometric graph (RGG) is constructed by placing points (nodes) according to a Poisson point process with density $\rho$ in a domain
${\cal V}\subseteq\mathbb{R}^d$, and linking
pairs of nodes with mutual distance less than $r_0$~\cite{Gilbert61}.    It remains a very
important model of spatial networks~\cite{Bart11}, where physical location of the nodes is important, for example climate~\cite{DZMK09},
infrastructure~\cite{RBJF15}, transport~\cite{BV08}, neuronal~\cite{BS12,NVSLB13} networks.  Perhaps surprisingly, it has also been shown
to be relevant to protein-protein interaction networks~\cite{HRP08}.    Many graph properties have been studied; here we focus on the property
of being connected, the existence of a multihop path between each pair of nodes.  We will sometimes use the synonymous term ``fully connected''
for consistency with previous literature.

RGGs are also increasingly being used to model wireless networks~\cite{HABDF09}, with focus on continuum percolation thresholds~\cite{SH13}
and clustering coefficients~\cite{DGJMN09}.
In the context of wireless ad-hoc networks, the nodes are devices that communicate directly with each other rather than via
a central router and whose locations are not specified in advance. The edges represent the ability of a pair of nodes
to communicate effectively. Percolation and connectivity thresholds for such models have previously been used to derive, for instance, the capacity of wireless
 networks~\cite{FDTT07}. Ad-hoc networks have many applications~\cite{CA11}, for example smart grid implementations, environmental monitoring, disaster
relief and emerging technologies such as the Internet of Things.

 Theoretical properties of RGGs have been widely studied by probabilists and combinatorialists~\cite{Walters11}.
A sequence of RGG is often considered, in which $\rho$, $r_0$ and the system size $L$ are varied at a specified rate such that the average number of nodes
$\mbox{\em const}\times\rho L^d\to\infty$.  Scaling all lengths (and hence these parameters) it is possible to fix any one of these quantities without loss of generality.
Here we fix $r_0$; for a discussion of limits with fixed $\rho$ or $L$ see Ref.~\cite{MA12}. Thus the following statements are made ``with high probability'' (whp),
meaning with a probability tending to unity in the combined limit.    At low densities (relative to $r_0$), the network consists of small clusters (connected components).
Beyond the percolation transition, the largest cluster becomes a macroscopic fraction of the size of the system.  If the domain is suitably well-behaved and $L$ is not
growing too rapidly, there is a further connectivity transition at which the graph forms a single cluster.  The latter may be described by $P_{fc}$, the probability of
(full) connectivity, which is a function of the density and the shape and size of the domain. 

The scaling for the connectivity transition that fixes $P_{fc}$ makes $L$ grow roughly exponentially with $\rho$.  For this scaling, the connection probability is dominated by
isolated nodes in the bulk (that is, far from the boundary)
for $d=2$ and near a two dimensional surface in $d=3$~\cite{Walters11}.  That is, in $d=2$ the larger number of nodes in the bulk dominates the
lower probability of links
for nodes near the boundary.  However, the present authors~\cite{CDG12b} have pointed out that for practical purposes, namely
approximating $P_{fc}$ in a realistic system, the size is not exponentially large, and either the bulk, edges or corners may dominate the connection probability~\cite{CDG12b}
depending on the density.  Thus, we are interested in results involving more general limiting processes, as well as useful approximations for finite cases.

Also motivated by the wireless applications, RGGs have been extended to a ``random connection model''~\cite{HABDF09,MA13,MA14},
also called ``soft RGG''~\cite{Penrose13}, in which pairs of nodes are linked
with independent probabilities $H(r)$ where $H$ typically decreases smoothly from $1$ to $0$
as the mutual distance $r$ increases from $0$ to $\infty$ (more general functions will be considered; see Sec.~\ref{s:H(r)}).
Thus there are two sources of randomness, the node locations and their links.
There are, however, a number of qualitative differences in connectivity between the hard and soft
connection models, for example, soft connections permit minimum degree as an effective proxy for k-connectivity~\cite{GDC14}.

The present authors have developed a theory to approximate $P_{fc}$ for soft connection functions and finite densities, expressing it as a sum of boundary
contributions~\cite{CDG12b,CDG12a}.  This can also be extended to anisotropic connection functions~\cite{CD13,GDC13b} to k-connectivity~\cite{GDC13a}
and to nonconvex domains~\cite{BDCR12,GDC13c}.

\subsection{Summary of new results}
The purpose of the present work is to upgrade this theory, increasing the generality and reducing cumbersome
calculations and uncontrolled approximations.  We start from the following approximate expression for $P_{fc}$ for $d\geq 2$, which effectively states that the
dominant contribution to lack of connectivity is that of isolated nodes, independently (Poisson) distributed
\begin{equation}\label{e:fc}
P_{fc}=\exp\left[-\rho\int_{\cal V} e^{-\rho M({\bf r})}d{\bf r}\right]
\end{equation}
with
\begin{equation}\label{e:M}
M({\bf r})=\int_{\cal V} H(|{\bf r-r'}|)d{\bf r}'
\end{equation}
the position-dependent connectivity mass.   In the case of $k$-connectivity we also need the
related integrals~\cite{GDC13a}
\begin{equation}
\int_{\cal V} M({\bf r})^m e^{-\rho M({\bf r})}d{\bf r}=(-1)^m\frac{d^m}{d\rho^m}\int_{\cal V} e^{-\rho M({\bf r})}d{\bf r}
\end{equation}
The integrals for connectivity and $k$-connectivity are four or six dimensional for $d=2,3$ respectively, and are almost never analytically tractable. 

Conditions under which Eq.~(\ref{e:fc}) is known rigorously (in the limit) are given in Ref.~\cite{Penrose13}; see also Ref.~\cite{Penrose15}.
Results are given for both Poisson and Binomial point
processes (the latter fixing the total number of nodes $N$ rather than the density $\rho$), including justifying the connection between connectivity and isolated
nodes for a class of connection functions of compact support, and the Poisson distribution of isolated nodes in a more general class that includes connection
functions that decay monotonically and at least exponentially fast at infinity.   However it is expected that most results and the above formula should be
valid more generally.   One exception is $d=1$ as isolated nodes are less relevant as the network may more readily split into two or more large pieces; the study
of this system for soft connection models remains an interesting open problem.  In practical situations we may be interested in $P_{fc}$ very close to unity; 
some literature approximates the exponential accordingly: $\exp(-z)\approx 1-z$.

The connection (and hence k-connection) probability $P_{fc}$ can then be written in ``semi-general'' form~\cite{CDG12b} as a sum of contributions from
different boundary elements.
\begin{equation}\label{e:SGF}
P_{fc}=\exp\left[-\sum_i \sum_{b\in{\cal B}_i} \rho^{1-i} G_{d,i}^{(b)} V_b e^{-\rho\Omega_b H_{d-1}}\right]
\end{equation}
where $0\leq i\leq d$ is the codimension of a boundary component $b$, $G_{d,i}^{(b)}$ is a geometrical factor obtained by expanding Eq.~(\ref{e:fc}) in the
vicinity of the boundary component, $V_b$ is the ($d-i$ dimensional) volume of the component $b$ (eg volume, surface area or edge length for $d=3$), $\Omega_b$
the magnitude of the available angular region, that is, its (solid) angle, and $H_{d-1}$ is a moment of the connection function, defined in Eq.~(\ref{e:Hm}) below.
To illustrate the notation, we give the case of a square domain:
\begin{equation}
P_{fc}=\exp\left[-\rho L^2 e^{-2\pi\rho H_1}-\frac{2L}{H_0}e^{-\pi\rho H_1}-\frac{4}{\rho H_0^2}e^{-\pi\rho H_1/2}\right]
\end{equation}
where the first term corresponds to the bulk, the second to the edges and the last to the corners.  If we specialise further to $H(r)=e^{-r^2}$,
the case of Rayleigh fading with $\eta=2$ and $\beta=1$ (see Tabs~\ref{t:ex} and~\ref{t:eta=2} below), the relevant moments are
$H_0=\sqrt{\pi}/2$, and $H_1=1/2$, and it becomes
\begin{equation}
P_{fc}=\exp\left[-\rho L^2 e^{-\pi\rho}-\frac{4L}{\sqrt{\pi}}e^{-\pi\rho/2}-\frac{16}{\pi\rho}e^{-\pi\rho/4}\right]
\end{equation}

This is now an explicit analytic expression, of much more
practical utility than Eq.~(\ref{e:fc}).

In previous work these contributions were computed separately for each connection function $H(r)$ by an asymptotic approximation to the integrals involving a number
of uncontrolled approximations.  Here, we provide the following improvements:
\begin{itemize}
\item Deriving these expansions for much more general connection functions, including all those commonly considered in the literature, allowing nonanalytic behaviour
at the origin and/or discontinuities: See Sec.~\ref{s:H(r)}.
\item Showing that the geometrical factor can be expressed simply as moments of $H$: See Tab.~\ref{t:G}.
\item Justifying the separation into boundary components, and stating it in a precise limiting form: See Sec.~\ref{s:sep}.
\item Finding the subleading (lower density) corrections, thus giving more accurate results at high density and a quantitative estimate of the range of validity: See
Secs.~\ref{s:P2D}-\ref{s:nonleading}.
\item Deriving the effects due to curvature for general smooth geometries in two and three dimensions: See Sec.~\ref{s:curv}. 
\end{itemize}
It should be emphasized that the approximation methods presented herein, encapsulated by Eq. (\ref{e:SGF}), significantly reduce the complexity of numerically
calculating the $d$-dimensional nested integrals of Eq.~(\ref{e:fc}). This is particularly useful when $H(r)$ is some special function (e.g. the Marcum $Q$-function).
Moreover, the linear form of the exponent in Eq.~(\ref{e:SGF}) enables direct analysis and comparison of contributions to $P_{fc}$ due to separate boundary components.

Section~\ref{s:H(r)} reviews previously used connection functions and defines the class of functions we consider here.
Section~\ref{s:mass} states the (more general) conditions on the connection function and derives expressions for the connectivity mass near boundaries.
Section~\ref{s:prob} then derives corresponding expressions and clarifications for $P_{fc}$.
Section~\ref{s:curv} extends the above calculations to domains with curved boundaries.
Section~\ref{s:ex} gives examples, showing that the results agree
with previous literature, and giving numerical confirmation of newly studied connection functions.
Section~\ref{s:conc} concludes.  

\begin{table}
\centerline{
\begin{tabular}{|c|c|c|c|c|}\hline
$G_{d,i}^\omega$&$i=0$&$i=1$&$i=2$&$i=3$\\\hline
$d=2$&1&$\frac{1}{2H_0}$&$\frac{1}{H_0^2\sin\omega}$&\\
$d=3$&1&$\frac{1}{2\pi H_1}$&$\frac{1}{\pi^2H_1^2\sin\omega}$&$\frac{4}{\pi^2H_1^3\omega\sin\omega}$\\\hline
\end{tabular}}
\caption{Geometric factors appearing in Eq.~(\protect\ref{e:SGF}) in terms of the dimension $d$ and boundary codimension $i$,
as first derived in Sec.~\protect\ref{s:prob} below.
Where $\omega$ appears it is the angle of a 2D corner or right angled 3D corner.  $H_m$ is the corresponding moment of the
connection function, Eq.~(\protect\ref{e:Hm}).\label{t:G}}
\end{table}

\begin{table*}
\centerline{
\begin{tabular}{|c|c|c|c|}\hline
&$H(r)$&$\tilde{H}_{s-1}$&Small $r$ expansion of $H(r)$\\\hline
Soft disk&$\left\{\begin{array}{cc}a&r<r_0\\0&r>r_0\end{array}\right.$&$\frac{a}{s}r_0^s$&$a$\\
Soft annulus&$\left\{\begin{array}{cc}a&r_-<r<r_+\\0&\mbox{otherwise}\end{array}\right.$&$\frac{a}{s}(r_+^s-r_-^s)$&$0$\\
Quasi unit disk&$\left\{\begin{array}{cc}1&r<r_-\\\frac{r_+-r}{r_+-r_-}&r_-<r<r_+\\0&r>r_+\end{array}\right.$&
$\left\{\begin{array}{cc}\frac{r_+^{s+1}-r_-^{s+1}}{s(s+1)(r_+-r_-)}&s\neq -1\\-\frac{\ln(r_+/r_-)}{r_+-r_-}&s=-1\end{array}\right.$&$1$\\
\hline
Waxman&$a e^{-\beta r}$&$a\beta^{-s}\Gamma(s)$&$a-a\beta r+\frac{a\beta^2r^2}{2}+\ldots$\\
\hline
Rayleigh SISO&$e^{-\beta r^\eta}$&$\eta^{-1}\beta^{-\frac{s}{\eta}}\Gamma\left(\frac{s}{\eta}\right)$&
$1-\beta r^\eta+\frac{\beta^2}{2}r^{2\eta}\ldots$\\
SIMO/MISO $(1,n)$&$\frac{\Gamma(n,\beta r^\eta)}{\Gamma(n)}$&$\frac{\beta^{-\frac{s}{\eta}}}{s\Gamma(n)}\Gamma\left(\frac{n+s}{\eta}\right)$&
$1-\frac{\beta^n}{n\Gamma(n)}r^{n\eta}+\frac{\beta^{n+1}}{(n+1)\Gamma(n)}r^{(n+1)\eta}\ldots$\\
MIMO $(2,2)$&$e^{-\beta r^\eta}((\beta r^{\eta})^2+2-e^{-\beta r^\eta})$&$\begin{array}{c}\frac{\left[(2-2^{-\frac{s}{\eta}})\eta^2+s\eta+s^2\right]}{\eta^{3}\beta^{s/\eta}}\Gamma\left(\frac{s}{\eta}\right)\end{array}$&
$1-\frac{\beta^4}{12}r^{4\eta}+\frac{\beta^5}{12}r^{5\eta}\dots$\\\hline
Rician&$Q_1(\sqrt{2K},\sqrt{2(K+1)\beta r^\eta})$&$\begin{array}{c}\frac{Q_{\frac{2s}{\eta}+1,0}(\sqrt{2K},0)}{s[2(K+1)\beta]^{s/\eta}}\end{array}$&
$1-\frac{(K-1)\beta}{e^K} r^\eta-\frac{(K+1)^2(K-1)\beta^2}{2e^K}r^{2\eta}\dots$\\
\hline
Log-normal&$\frac{1}{2}\mbox{erfc}\left(\frac{10}{\sigma\sqrt{2}}\log_{10}(\beta r^\eta)-\frac{\mu}{\sigma\sqrt{2}}\right)$&
$\frac{10^{\frac{\mu s}{10\eta}}e^{\frac{1}{2}\left(\frac{s\sigma\ln 10}{10\eta}\right)^2}}{s\beta^{s/\eta}}$&
$1-\frac{1}{2\sqrt{\pi}}e^{\left(\frac{10}{\sqrt{2}\sigma}\log_{10}(\beta r^\eta)-\frac{\mu}{\sqrt{2}\sigma}\right)^2}+\ldots$\\
\hline
\end{tabular}}
\caption{Explicit examples discussed in Sec.~\protect\ref{s:lit}.
$\tilde{H}_m=H_m$ for $m\geq 0$, Eq.~(\protect\ref{e:Hm}); otherwise  the relevant generalisation, Eq.~(\protect\ref{e:H-2}). 
$Q_m$ is the Marcum Q-function and $Q_{m,n}$ its generalisation, the Nuttall Q-function~\cite{Nuttall72,Simon02}. \label{t:ex}}
\end{table*}

\section{General connection functions}\label{s:H(r)}
\subsection{Connection functions appearing in the wireless communications literature}\label{s:lit}
The connection function $H(r)$ gives the probability of a direct link between two nodes at distance $r$.  We want to construct a class of connection functions 
$H:[0,\infty)\to[0,1]$ that includes virtually all of those appearing in the existing literature; refer to Tab.~\ref{t:ex}.  While the most developed models have
appeared in the wireless communications literature,  it is not difficult to measure and model the distance-dependent link
probability in other spatial networks; see for example Ref.~\cite{NVSLB13}.

The original Gilbert random geometric graph (``unit disk'' or ``hard disk'') model~\cite{Gilbert61}, considered in most of the subsequent literature~\cite{Walters11},
is deterministic - all links
are made within a fixed pairwise distance $r_0$ and none otherwise.  In Tab.~\ref{t:ex} it 
is the soft disk with $a=1$.  The soft disk itself was considered by Penrose~\cite{Penrose13}, who noted that its edge set corresponds to the intersection of
those of the Gilbert and Erd\H{o}s-R\'enyi (fixed probability for links) random graph models.  A (deterministic) annulus has also been considered~\cite{BBW04}.
Such models may be of interest when dealing with encrypted messages of packet forwarding networks where communication links should only form with
distant neighbours as to avoid interference or a security breach. A quasi unit disk model~\cite{KWZ03} is one in which all links
are made within a range $r_-$ and none with range greater than $r_+$.
While this is sufficient to observe interesting phenomena and prove bounds, a specific model requires a method for determining (deterministically or probabilistically)
the links
lying between $r_-$ and $r_+$.  One natural such approach, given in Ref.~\cite{GCFN11}, gives an $H(r)$ decreasing linearly between
these points, as presented in Tab.~\ref{t:ex}.  In all these examples, the connection function is not a smooth function of distance, so our class of functions must
allow discontinuities in the function and/or its derivatives.  

Another main class of connection functions comes from fading models that take account of noise in the transmission channel but neglect interference from other
signals.  Interference is often of relevance but leads to models beyond the scope of this work~\cite{IRK09}; it may be mitigated by transmitting at different frequencies
and/or at different times.  The received signal is in general a combination of specular (coherent) and diffusive (incoherent) components~\cite{DRW02}.  The diffusive
component leads to the Rayleigh fading model of Tab.~\ref{t:ex}, while a combination of diffusive and a single specular component leads to the Rician model.  The parameter
$K$ controls the relative strength of these two components, so that the Rician model limits to Rayleigh as $K\to 0$.  Models with more than one specular component
lead to similar but more involved expressions, which can also be approximated using the same functions as in the Rician case, but with slightly different
parameters~\cite{DRW02,CD13}.

A further extension is to consider multiple antennas for transmission and reception (MIMO, ie multiple input and multiple output), or for one of these (MISO, multiple input
single output, or SIMO).  Combining Rayleigh channels with maximum ratio combining (MRC) at the receiver leads to the expressions given in Tab.~\ref{t:ex}; see
Refs.~\cite{BDCR12,CGD14,KA03}.  Note that SIMO/MISO reduces to the original (SISO) Rayleigh model when $n=1$; for real parameter $m=n\geq 1/2$ it
takes the same form as the Nakagami-$m$ fading model, of more general applicability and interest~\cite{AG00}.

Finally, slow fading, due to larger obstacles that do not move appreciably on the timescale of wireless transmission, is often modelled by the log-normal
distribution~\cite{AS02}.  This leads to a connection function which is smooth but has vanishing derivatives of all orders at the origin.  Note, however, that
the assumption of independence of the probability of each link may be more difficult to justify here.

\begin{table*}
\centerline{
\begin{tabular}{|c|c|c|c|c|}\hline
&$H_2$&$H_1$&$H_0$&$\tilde{H}_{-2}$\\\hline
SISO $(1,1)$&$\frac{1}{4}\sqrt{\frac{\pi}{\beta^3}}$&$\frac{1}{2\beta}$&$\frac{1}{2}\sqrt{\frac{\pi}{\beta}}$&$-\sqrt{\pi\beta}$\\
MIMO $(2,2)$&$\frac{23-\sqrt{2}}{16}\sqrt{\frac{\pi}{\beta^3}}$&$\frac{7}{4\beta}$&$\frac{11-2\sqrt{2}}{8}\sqrt{\frac{\pi}{\beta}}$&$\frac{4\sqrt{2}-7}{4}\sqrt{\pi\beta}$\\\hline
Rician&$\begin{array}{c}\sqrt{\frac{\pi}{(K+1)^3\beta^3}}e^{-\frac{K}{2}}\\\times\left[\frac{2K^2+6K+3}{12}I_0(\frac{K}{2})+\frac{K^2+2K}{6}I_1(\frac{K}{2})\right]\end{array}$
&$\frac{1}{2\beta}$&$\begin{array}{c}\sqrt{\frac{\pi}{(K+1)\beta}}e^{-\frac{K}{2}}\\\times\left[\frac{K+1}{2}I_0(\frac{K}{2})+\frac{K}{2}I_1(\frac{K}{2})\right]\end{array}$&
$\begin{array}{c}-\sqrt{(K+1)\pi\beta}e^{-\frac{K}{2}}\\\times I_0(\frac{K}{2})\end{array}$\\
\hline
\end{tabular}}
\caption{Examples from Tab.~\protect\ref{t:ex} for $\eta=2$ and specific $s$. $I_m$ denotes a modified Bessel function.\label{t:eta=2}}
\end{table*}

In all of these models, the expression $r^2$ appears naturally, coming from the inverse square law for signal intensity in three dimensional
space, see Tab.~\ref{t:eta=2}.  However many authors consider a more general $r^\eta$, with the path loss exponent~\cite{EGTPGKJB99} $\eta$ varying from
$1$ (signal strictly confined to a two dimensional domain with no absorption) to about $6$ (more cluttered/absorptive environments).  The path loss exponent may also be
used to interpolate between random and deterministic models, for example the Rayleigh fading function $\exp[-(r/r_0)^\eta]$ tends to the unit disk model as $\eta\to\infty$.
The inclusion of non-integer $\eta$ requires us to allow series expansions of $H(r)$ with non-integer powers at the origin.

Normally in ad-hoc networks the path loss exponent is significantly greater than unity.  However, Waxman~\cite{Waxman88} developed a model with $\eta=1$ and in
general a nontrivial coefficient in front of the exponential for more general large networks.  Zegura et al~\cite{ZCA96} use this as a model of the internet, and also propose
the connection function $H(r)=\alpha\exp[-r/(L-r)]$, however long range links
proportional to the system size are beyond the scope of our approximations.

Some works add a small length scale to avoid an unphysical divergent signal strength at the origin, for example replacing $r^\eta$ by $r^\eta+\varepsilon$.  For an explicit
number of transmitters $n$ it is straightforward to perform the integrals in the case of Rayleigh fading SISO/MISO/MIMO, but for reasons of clarity have been omitted from
Tab.~\ref{t:ex}.

\begin{figure*}[t]
\centerline{\includegraphics[width=500pt]{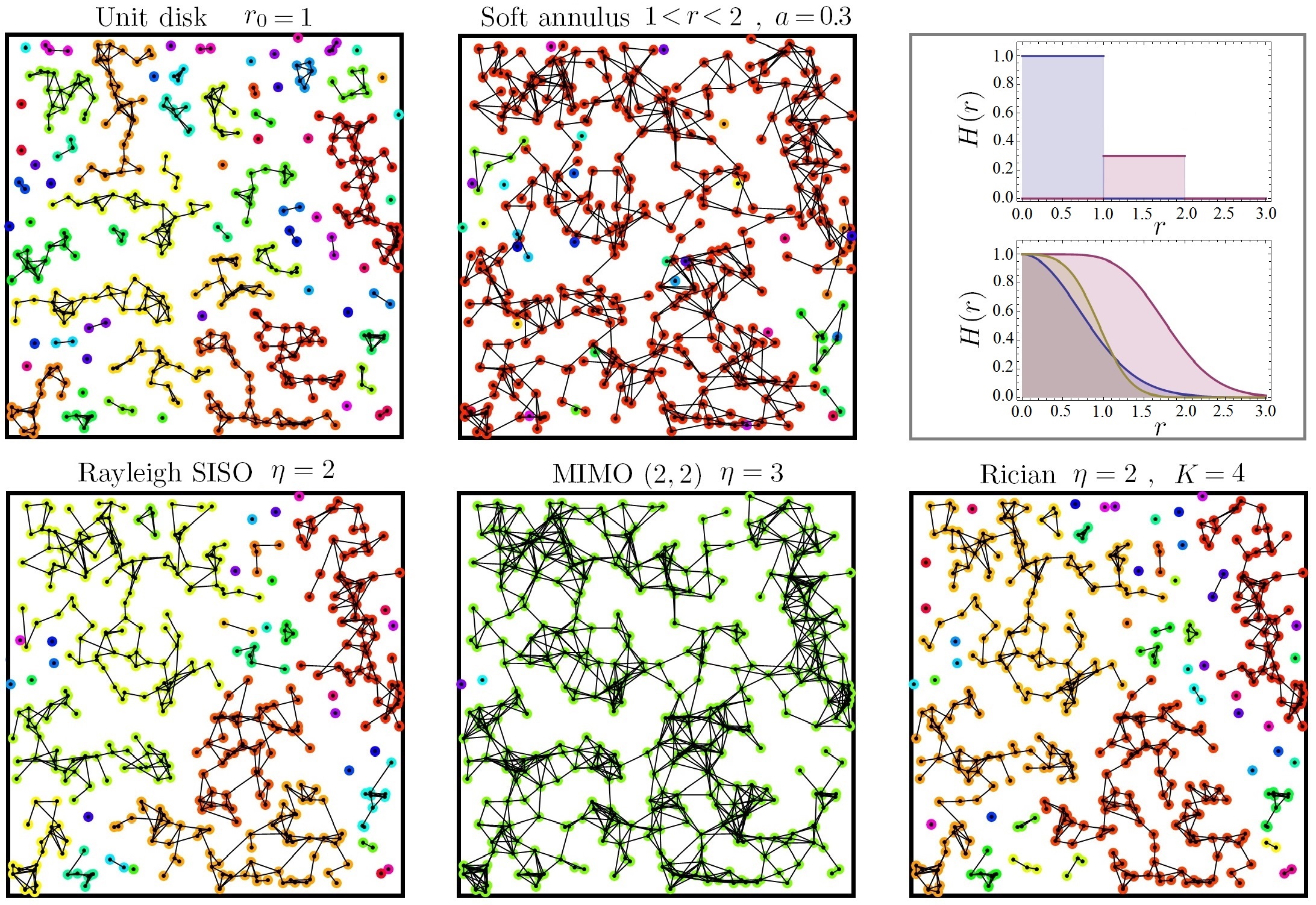}}
\caption{Networks of $N=400$ nodes in a square domain of side $L=20$ resulting from five different connection functions using the same random seed. Clusters of nodes are colour coded. In the top right panel we plot these five $H(r)$ functions in the respective panels. In the lower three we use $\beta=1$.\label{f:nets}}
\end{figure*}

Fig.~\ref{f:nets} shows the effects of several connection functions in forming a RGG; note the striking differences in network topologies.
In the simulations, spatial coordinates for $N$ nodes are chosen at random inside a square
domain. Nodes $i$ and $j$ are then paired with independent probabilities $H(r_{ij})$.  The resulting links
are stored in a symmetric zero-one adjacency matrix,
and a depth-first search algorithm identifies the connected components of the graph, a process of complexity order $\mathcal{O}(N \ln N)$. For Fig.~\ref{f:nets} we use
the same random seed as to allow comparison between the different $H(r)$ functions as plotted in the top right panel of Fig.~\ref{f:nets}, however the process can be
repeated in a Monte Carlo fashion (with random seeds) and for different values of $\rho=N/V$ to generate Fig.~\ref{f:R2} below.

\subsection{Assumptions and notation}\label{s:ass}
Based on these existing examples, we make the following assumptions:
\begin{enumerate}
\item Near the origin, $H(r)$ is described by the expansion
\begin{equation}
H(r)=H(0)+\sum_{\alpha\in{\cal A}} a_\alpha r^{\alpha}
\end{equation}
where ${\cal A}\subset (0,\infty)$ has a positive lower bound on the gap size.  The minimum of $\cal A$ is denoted
$\alpha_{min}$.
\item $H(r)$ is piecewise smooth, with non-smooth points
at a discrete and possibly empty set $\{r_k\}$, $k\in{\cal K}$, also with a positive lower bound on the gap size.
\item The bulk connectivity mass
\begin{equation}
M_{d,0}=\left\{\begin{array}{cc}2\pi\int H(r)rdr&d=2\\4\pi\int H(r)r^2dr&d=3\end{array}\right.
\end{equation}
 is finite.
\item All derivatives of $H(r)$ are monotonic for sufficiently large $r$.
\end{enumerate}
Remarks: We have $H(0)\in[0,1]$ in all cases.   In the case of log-normal fading, corrections for small
$r$ are smaller than any power of $r$, ie the expansion is just $H(0)$.  If the connectivity mass is
finite but $H(r)$ decays very slowly at infinity, some of the local assumptions (and hence
Eq.~\ref{e:fc}) may
fail; Mao and Anderson~\cite{MA12} insist on $H(r)=o(r^{-2}\ln(r)^{-2})$ at infinity,
which is slightly stronger than finite connectivity mass in $d=2$, for some
of their results, but we are mostly interested in exponential decay.  The
final assumption is to ensure sufficiently rapid decay of the derivatives of
$H(r)$ at infinity.

The function $H(r)$ describes the link
probability on a line passing through a particular node;
we will need various moments of this:
\begin{eqnarray}
H_m&=&\int_0^\infty H(r) r^mdr \nonumber\\
H_m'&=&\int_0^\infty H'(r) r^mdr \label{e:Hm}\\
H_m''&=&\int_0^\infty H''(r) r^m dr \nonumber
\end{eqnarray}
Integration by parts gives $H_m'=-mH_{m-1}$ for $m>0$ and $H_m''=m(m-1)H_{m-2}$ for $m>1$,
however the form of $H(r)$ implies that $H_m'$ and $H_m''$ have a greater range of validity, since
the constant $H(0)$ is removed by differentiation:
Our assumptions imply that $H_m$ is defined for $-1<m\leq d-1$, $H_m'$ is defined
for $-\alpha_{min}<m\leq d$  and $H_m''$ is defined for $1-\alpha_{min}<m\leq d+1$.  Where there is
an explicit formula for non-integer $H_m$ and hence for sufficiently large $m$ for $H_m'$ and
$H_m''$, it may be used to analytically continue the expression to lower $m$; examples are given
in Sec.~\ref{s:ex} below.

The moments may be considered as a Mellin transform evaluated at particular values which depend on
the path loss exponent $\eta$.  So, if we have $H(r)=g(\beta r^\eta)$ for some scaled function $g$, a straightforward
change of variables gives
\begin{equation}
H_{s-1}=\frac{1}{\eta\beta^{s/\eta}}\{{\cal M}g\}\left(\frac{s}{\eta}\right)
\end{equation}
where
\begin{equation}
\{{\cal M} g\}(u)=\int_0^\infty t^{u-1} g(t) dt 
\end{equation} 
is the Mellin transform of the function $g$.

Occasionally we also define incomplete versions of the moments
\begin{equation}
H_m(\epsilon)=\int_\epsilon^\infty H(r)r^mdr
\end{equation}
and similarly for the primed versions.

We will also need to define contributions from discontinuities,
\begin{eqnarray}
\Delta_m&=&\sum_{k\in{\cal K}}r_k^m\left[H(r_k+)-H(r_k-)\right]\\
\Delta_m'&=&\sum_{k\in{\cal K}}r_k^m\left[H'(r_k+)-H'(r_k-)\right]
\end{eqnarray}

For the most general calculations we use the further notation
\begin{widetext}
\begin{eqnarray}
\tilde{H}_{-2}&=&\left\{\begin{array}{lc}H_{-1}'+\Delta_{-1}&\alpha_{min}>1\\
\displaystyle\lim_{\epsilon\to 0}\left(H_{-1}'(\epsilon)+a_1\ln\left(|a_1|\epsilon\right)+\displaystyle\sum_{\alpha\neq 1}a_\alpha\frac{\alpha\epsilon^{\alpha-1}}{\alpha-1}\right)+\Delta_{-1}
&\alpha_{min}\leq 1\end{array}\right.\label{e:H-2}\\
3\tilde{H}_{-4}&=&\left\{\begin{array}{lc}H_{-2}''-H_{-3}'+\Delta_{-2}'+\Delta_{-3}&\alpha_{min}>3\\&\\
\displaystyle\lim_{\epsilon\to0}\left(H_{-2}''(\epsilon)-H_{-3}'(\epsilon)+3a_3\ln\left(|a_3|^{1/3}\epsilon\right)
+\displaystyle\sum_{\alpha\neq 3}a_\alpha\frac{\alpha(\alpha-2)\epsilon^{\alpha-3}}{(\alpha-3)}\right)+\Delta_{-2}'+\Delta_{-3}
&\alpha_{min}\leq 3\end{array}\right.\nonumber
\end{eqnarray}
\end{widetext}
It turns out, however, that in most cases we need the expansions only to second order in the small parameter, and can also assume $\alpha_{min}>1$ (except for
the Waxman model).
In this case most of the technical details can be avoided, and we find that $\tilde{H}_{-2}$ is given only by the first option, $\tilde{H}_{-4}$ is not needed at all,
and terms involving $a_\alpha$ (including the hypergeometric functions below) are also not required.  So, the reader can safely omit these terms at first sight,
and consider them only when a fuller and more general understanding is required.

\begin{figure}
\centerline{\includegraphics[width=400pt]{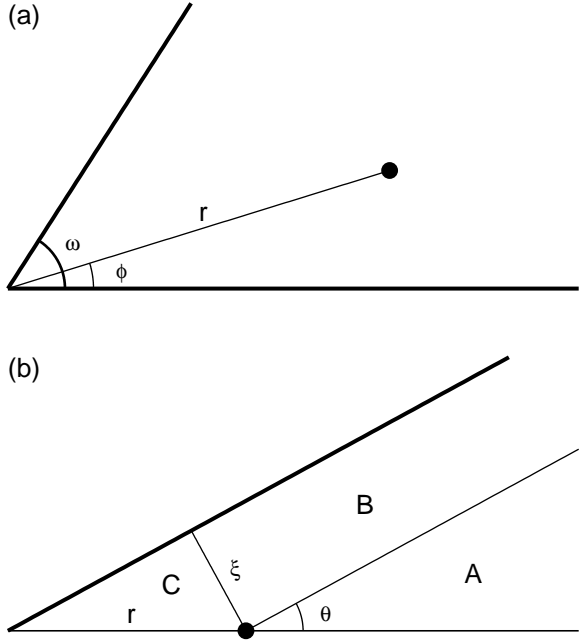}}
\caption{(a) Geometry for the connectivity mass of a wedge, with the node at the indicated point.  We split the domain into two wedges, each with the node
located at the edge.  (b) The resulting wedge is further split into three parts.\label{f:wedge}}
\end{figure}

\section{Connectivity mass}\label{s:mass}
\subsection{Integration on a non-centred line}
The computation of connection probability, Eq.~(\ref{e:fc}) for moderate to large density is dominated by contributions from the bulk and various boundary
components.  Each boundary component is controlled by the form of the connectivity mass at and near the boundary, the calculation to which we turn first.
This  subsection deals with the first integral of $H(r)$, that on an off-centre line, that is needed for the calculations in the later subsections, on the
2D and 3D connectivity mass, respectively.

The connection function is first integrated on a line passing a small distance $x>0$ from the node:
\begin{equation}
F(x)=\int_0^\infty H(\sqrt{x^2+t^2})dt
\end{equation}
If $\alpha_{min}>3$ we can expand for small $x$ to get
\begin{eqnarray}
F(x)&=&H_0+\frac{x^2}{2}[H_{-1}'+\Delta_{-1}]\label{e:Fexp}\\
&&+\frac{x^4}{8}[H_{-2}''-H_{-3}'+\Delta_{-2}'+\Delta_{-3}]+o(x^4)\nonumber
\end{eqnarray}
This may be derived by splitting the integral at the discontinuities, differentiating the result
(including integrand and limits) with respect to $x^2$ to get the coefficients of the Taylor series.

If $\alpha_{min}\leq 3$ the integrals $H_{-3}'$ and $H_{-2}''$ diverge, and if $\alpha_{min}\leq 1$ the integral $H_{-1}'$
also diverges.  In this case we need to split the integrals at a point $\epsilon\ll 1$, and use the small $r$ expansion
of $H(r)$ to treat the contribution near the origin separately.  We require that $\epsilon$ is much larger than any positive
power of $x$; formally we take the limit $x\to 0$ and only then $\epsilon\to 0$.  By analogy with the incomplete
gamma function, we denote
\begin{eqnarray}
f(x,\epsilon)&=&\int_0^\epsilon H(\sqrt{x^2+t^2})dt\\
F(x,\epsilon)&=&\int_\epsilon^\infty  H(\sqrt{x^2+t^2})dt
\end{eqnarray}

For $f(x,\epsilon)$,  make a change of variable:
\begin{eqnarray}\label{e:s}
f(x,\epsilon)&=&\int_1^{\sqrt{(\epsilon^2/x^2)+1}} H(xs) \frac{xs\;ds}{\sqrt{s^2-1}}\\
s&=&\sqrt{(t^2/x^2)+1}\nonumber
\end{eqnarray}
Then, expanding for small $x$ (at fixed $\epsilon$) we have
\begin{eqnarray}
f(x,\epsilon)&=&H(0)\epsilon+\sum_{\alpha} a_\alpha f_{\alpha}(x,\epsilon)\\
f_{\alpha}(x,\epsilon)&=&\frac{\sqrt{\pi}}{2}\frac{\Gamma(-\frac{\alpha+1}{2})}{\Gamma(-\frac{\alpha}{2})}x^{\alpha+1}
+\frac{\epsilon^{\alpha+1}}{\alpha+1}+\frac{\alpha\epsilon^{\alpha-1}}{2(\alpha-1)}x^2\nonumber\\
&&+\frac{\alpha(\alpha-2)\epsilon^{\alpha-3}}{8(\alpha-3)}x^4+o(x^4)\nonumber
\end{eqnarray}
except that if any of the $\alpha$ are odd integers two of the terms diverge and are replaced by a logarithm:
\begin{eqnarray*}
f_1(x,\epsilon)&=&\frac{\epsilon^2}{2}+\frac12\ln\left(\frac{2\sqrt{e}\epsilon}{x}\right)x^2+\frac{x^4}{16\epsilon^2}+o(x^4)\\
f_3(x,\epsilon)&=&\frac{\epsilon^4}{4}+\frac{3\epsilon^2}{4}x^2+\frac{3}{8}\ln\left(\frac{2e^{3/4}\epsilon}{x}\right)x^4+o(x^4)
\end{eqnarray*}
For even integers (for example the well-studied case $H(r)=e^{-r^2}$), the $x^{\alpha+1}$ term is zero, and the series is finite.

The upper integral $F(x,\epsilon)$ has the same expansion as Eq.~(\ref{e:Fexp}), but with incomplete moments $H_m(\epsilon)$, $H_m'(\epsilon)$ and $H_m''(\epsilon)$.

Putting it back together, we have
\begin{widetext}
\begin{equation}
F(x)=H(0)\epsilon+\sum_\alpha a_\alpha f_{\alpha}(x,\epsilon)+H_0(\epsilon)+\frac{x^2}{2}[H_{-1}'(\epsilon)+\Delta_{-1}]
+\frac{x^4}{8}[H_{-2}''(\epsilon)-H_{-3}'(\epsilon)+\Delta_{-2}'+\Delta_{-3}]+o(x^4)
\end{equation}
Note that $F(x)$ does not depend on $\epsilon$: All $\epsilon$ where the relevant series converge should be equivalent, and in particular we may set $\epsilon=0$ where
possible to reconstitute the full moment, and otherwise take the limit of a regularised version.  So, collecting terms by powers of $x$ we have finally
\begin{equation}
F(x)=H_0+\frac{x^2}{2}\left[\tilde{H}_{-2}+a_1\ln\left(\frac{2\sqrt{e}}{|a_1|x}\right)\right]+\frac{3x^4}{8}\left[\tilde{H}_{-4}+a_3\ln\left(\frac{2e^{3/4}}{|a_3|^{1/3}x}\right)\right]+
\sum_{\alpha\not\in\{1,3\}}a_\alpha\frac{\sqrt{\pi}}{2}\frac{\Gamma(-\frac{\alpha+1}{2})}{\Gamma(-\frac{\alpha}{2})}x^{\alpha+1}+o(x^4)
\end{equation}
\end{widetext}
where $a_1$ and/or $a_3$ are deemed to be zero if they do not appear in the expansion of $H(r)$.  If they do appear, $|a_1|$ and $|a_3|^{1/3}$ are
included to ensure that the argument of each logarithm is dimensionless.  An $a_2$ term contributes only at order $x^2$:
Both $x^3$ and $x^4$ coefficients vanish.  Note that if there are no discontinuities, $\tilde{H}_{-2}$ and $\tilde{H}_{-4}$
correspond to the continuation of the integration by parts expression of $H_{-2}$ and $H_{-4}$ (respectively) to negative index.

\subsection{Connectivity mass of polygons}\label{s:wedge}
Here we find expansions for the connectivity mass defined in Eq.~(\ref{e:M}) on and near the boundary.
We will use $M_{d,i}$ to denote the mass near a boundary
where $d$ is the dimension and $i$ the boundary codimension.  The dependence of $M_{d,i}$ on variables may be implicit in the notation;  in general it may depend
on a parameter (for example the wedge angle) as well as the node location in an appropriate coordinate system.

This section deals with $d=2$, while the next deals with $d=3$.
We consider a wedge of total angle $\omega$ and node position in polar coordinates $(r,\phi)$ with connectivity mass denoted
$M_{2,2}^\omega(r,\phi)$. Using first the simplified geometry of Fig.~\ref{f:wedge} with $\xi$ a small parameter and (for now) the node on the boundary,
the connectivity mass is the sum of three contributions A, B and C as follows:
\begin{eqnarray}
M_{2,2}^\theta(\xi\csc\theta,0)&=&M_{2A}+M_{2B}+M_{2C}\\
M_{2A}&=&\int_0^\theta d\phi\int_0^\infty H(r)rdr=\theta H_1\\
M_{2B}&=&\int_0^\xi dx\int_0^\infty dt H(\sqrt{x^2+t^2})\nonumber\\
&=&\int_0^\xi dx F(x)\\
M_{2C}&=&\int_0^\xi dx\int_0^{x\cot\theta} dt H(\sqrt{x^2+t^2})\nonumber\\
&=&\int_0^\xi dx f(x,x\cot\theta)
\end{eqnarray}
where the $2$ in $2A$ etc. denotes the dimension.
$M_{2B}$ may be found by integrating the expressions in the previous section, noting that $\xi$, and hence $x$, is small.  For $M_{2C}$, we have $x$ small,
but we do not have $x\cot\theta\gg x$, so the previous separation between $\epsilon$ and $x$ does not apply. Instead, integrate  Eq.~(\ref{e:s}) directly to obtain
$f(x,x\cot\theta)$ and hence
\begin{widetext}
\begin{eqnarray}
M_{2C}&=&\sum_{\alpha\not\in\{1,3\}}a_\alpha\xi^{\alpha+2}\left[
\frac{\sqrt{\pi}\Gamma(\frac{-1-\alpha}{2})}{2(\alpha+2)\Gamma(\frac{-\alpha}{2})}+\right.
\left.\frac{\csc^{\alpha+1}\theta}{(\alpha+1)(\alpha+2)}\;\:F_{\hspace{-10pt}2\hspace{5pt}1}\left(\frac{1}{2},\frac{-1-\alpha}{2},\frac{1-\alpha}{2},\sin^2\theta\right)\right]\\
&&+\frac{H(0)\xi^2}{2}\frac{\cos\theta}{\sin\theta}+\frac{a_1\xi^3}{6}\left[\frac{\cos\theta}{\sin^2\theta}+\ln\frac{1+\cos\theta}{\sin\theta}\right]
+\frac{a_3\xi^5}{40}\left[\frac{\cos\theta}{\sin^2\theta}\left(3+\frac{2}{\sin^2\theta}\right)+3\ln\frac{1+\cos\theta}{\sin\theta}\right]+o(\xi^5)\nonumber
\end{eqnarray}
Thus, for the whole wedge we have
\begin{eqnarray}
M_{2,2}^\theta(\xi\csc\theta,0)&=&\theta H_1+\xi H_0+\xi^2\frac{H(0)}{2}\frac{\cos\theta}{\sin\theta}+\frac{\xi^3}{6}\left[\tilde{H}_{-2}+
a_1\left(\frac{\cos\theta}{\sin^2\theta}+\ln\frac{1+\cos\theta}{\sin\theta}+\ln\left(\frac{2e^{5/6}}{|a_1|\xi}\right)\right)\right]\nonumber\\\label{e:wholewedge}
&&+\frac{\xi^5}{40}\left[3\tilde{H}_{-4}+a_3\left(\frac{\cos\theta}{\sin^2\theta}
\left(3+\frac{2}{\sin^2\theta}\right)+3\ln\frac{1+\cos\theta}{\sin\theta}+3\ln\left(\frac{2e^{19/20}}{|a_3|^{1/3}\xi}\right)\right)\right]\\
&&+\sum_{\alpha\not\in\{1,3\}}a_\alpha\xi^{\alpha+2}\left[
\frac{\sqrt{\pi}\Gamma(\frac{-1-\alpha}{2})}{2(\alpha+2)\Gamma(\frac{-\alpha}{2})}+\right.\nonumber
\left.\frac{\csc^{\alpha+1}\theta}{(\alpha+1)(\alpha+2)}\;\:F_{\hspace{-10pt}2\hspace{5pt}1}\left(\frac{1}{2},\frac{-1-\alpha}{2},\frac{1-\alpha}{2},\sin^2\theta\right)\right]
+o(\xi^5)
\end{eqnarray}

There are special values of the hypergeometric function for even $\alpha$:
\begin{eqnarray}
\left.F_{\hspace{-10pt}2\hspace{5pt}1}\left(\frac{1}{2},\frac{-1-\alpha}{2},\frac{1-\alpha}{2},\sin^2\theta\right)\right|_{\alpha=2}&=&\cos\theta(3-\cos^2\theta)\\
\left.F_{\hspace{-10pt}2\hspace{5pt}1}\left(\frac{1}{2},\frac{-1-\alpha}{2},\frac{1-\alpha}{2},\sin^2\theta\right)\right|_{\alpha=4}&=&\frac{1}{3}
\cos\theta(15-20\cos^2\theta+8\cos^4\theta)
\end{eqnarray}
and for limiting angles
\begin{eqnarray}
\left.F_{\hspace{-10pt}2\hspace{5pt}1}\left(\frac{1}{2},\frac{-1-\alpha}{2},\frac{1-\alpha}{2},\sin^2\theta\right)\right|_{\theta=0}
&=&\left.F_{\hspace{-10pt}2\hspace{5pt}1}\left(\frac{1}{2},\frac{-1-\alpha}{2},\frac{1-\alpha}{2},\sin^2\theta\right)\right|_{\theta=\pi}=1\\
\left.F_{\hspace{-10pt}2\hspace{5pt}1}\left(\frac{1}{2},\frac{-1-\alpha}{2},\frac{1-\alpha}{2},\sin^2\theta\right)\right|_{\theta=\frac{\pi}{2}}&=&
\frac{\sqrt{\pi}\Gamma(\frac{1-\alpha}{2})}{\Gamma(\frac{-\alpha}{2})}
\end{eqnarray}
Thus the two terms in the sum over $\alpha$ cancel when $\theta=\pi/2$.
 
Combining two wedges, we have the connectivity mass at a general point of a wedge of angle (and solid angle) $\omega$, with the node at polar point $(r,\theta)$:
\begin{eqnarray}
M_{2,2}^\omega(r,\theta)&=&M_{2,2}^\theta(r,0)+M_{2,2}^{\theta'}(r,0)\label{e:wedge}\\
&=&\omega H_1+rH_0[\sin\theta+\sin\theta']+\frac{H(0)r^2}{2}[\sin\theta\cos\theta+\sin\theta'\cos\theta']\nonumber\\
&&+\frac{\tilde{H}_{-2}r^3}{6}[\sin^3\theta+\sin^3\theta']+\frac{3\tilde{H}_{-4}r^5}{40}[\sin^5\theta+\sin^5\theta']+O(r^{\alpha_{min}+2}\ln r)+o(r^5)
\end{eqnarray}
where $\theta'=\omega-\theta$ and omitted terms involving $a_\alpha$ can be found from Eq.~(\ref{e:wholewedge}) above.

An important special case is that of an edge, $M_{2,1}$ where $\omega=\pi$ and we may take $\theta=\pi/2$, and so the $a_\alpha$ terms cancel.
The above expressions reduce to
\begin{eqnarray}
M_{2,1}(r)&=&\pi H_1+2rH_0+\frac{r^3}{3}\left[\tilde{H}_{-2}+a_1\ln\left(\frac{2e^{5/6}}{|a_1|r}\right)\right]+\frac{3r^5}{20}\left[\tilde{H}_{-4}+a_3\ln\left(\frac{2e^{19/20}}{|a_3|^{1/3}r}\right)\right]+o(r^5)
\end{eqnarray}

Together with the bulk connectivity mass $M_{2,0}=2\pi H_1$ we have all the ingredients for convex polygons.

\subsection{Connectivity mass of polyhedra}\label{s:3D}
Here, we find the connectivity mass near the boundary in three dimensional geometries.
The connectivity mass of the bulk is $M_{3,0}=4\pi H_2$.  For  a node a small distance $r$ from a face we use cylindrical coordinates and
a transformation $ s=\sqrt{x^2+\rho^2}$ in the second line:
\begin{eqnarray}
M_{3,1}(r)&=&\int_{-r}^\infty dx \int_0^\infty \rho \;d\rho \int_0^{2\pi} d\theta \;H(\sqrt{x^2+\rho^2}) \nonumber\\
&=&M_{3,1}(0)+2\pi\int_0^r dx \int_x^\infty ds \;sH(s)\\
&=&2\pi H_2+2\pi\int_0^r dx \left[H_1-\int_0^x ds \;sH(s)\right]\nonumber \\
&=&2\pi\left[H_2+rH_1-\frac{r^3H(0)}{6}-\sum_\alpha a_\alpha\frac{r^{3+\alpha}}{(\alpha+2)(\alpha+3)}\right]\nonumber
\end{eqnarray}

For an edge in 3D of angle $\theta$, use the same splitting and coordinates as in Fig.~\ref{f:wedge}, that is, first consider a node on the boundary,
then the interior case consists of two combined 3D wedges.  Noting that the solid angle is $2\theta$, we find
\begin{eqnarray}
M_{3,2}^\theta(\xi\csc\theta,0)&=&M_{3A}+M_{3B}+M_{3C}\\
M_{3A}&=&2\theta H_2\\
M_{3B}&=&\pi\left[\xi H_1-\frac{\xi^3H(0)}{6}-\sum_\alpha a_\alpha\frac{\xi^{3+\alpha}}{(\alpha+2)(\alpha+3)}\right]\label{e:3B}
\end{eqnarray}
Note that the $B$ region is half the slab considered for the face above.  Using cylindrical coordinates we have
\begin{eqnarray}
M_{3C}&=&\int_0^{\pi/2-\theta}\int_0^{\xi\sec\phi}\int_{-\infty}^{\infty}\rho H(\sqrt\rho^2+z^2)dzd\rho d\phi\nonumber\\
&=&2\int_0^{\pi/2-\theta}\int_0^{\xi\sec\phi}\rho F(\rho)d\rho d\phi\nonumber\\
&=&\int_0^{\pi/2-\theta}\left\{H_0\xi^2\sec^2\phi+\frac{\xi^4\sec^4\phi}{4}\left[\tilde{H}_{-2}+a_1\ln\left(\frac{2e^{3/4}}{|a_1|\xi\sec\phi}\right)\right]
+\sum_{\alpha\neq 1}\frac{a_\alpha\sqrt{\pi}}{3+\alpha}\frac{\Gamma(-\frac{\alpha+1}{2})}{\Gamma(-\frac{\alpha}{2})}\xi^{3+\alpha}\sec^{3+\alpha}\phi
\right\}d\phi\nonumber\\
&&+o(\xi^4)\nonumber\\
&=&H_0\xi^2\cot\theta+\frac{\xi^4}{4}(\cot\theta+\frac{\cot^3\theta}{3})\left[\tilde{H}_{-2}+a_1\ln\left(\frac{2e^{3/4}}{|a_1|\xi\csc\theta}\right)\right]\label{e:3C}\\
&&+\sum_{\alpha\neq 1}\frac{a_\alpha\sqrt{\pi}}{3+\alpha}
\frac{\Gamma(-\frac{\alpha+1}{2})}{\Gamma(-\frac{\alpha}{2})}\xi^{3+\alpha}\cos\theta\hspace{5pt}
F_{\hspace{-10pt}2\hspace{5pt}1}\left(\frac{1}{2},\frac{4+\alpha}{2},\frac{3}{2},\cos^2\theta\right)+o(\xi^4)\nonumber
\end{eqnarray}
Thus the combined edge contribution is
\begin{equation}
M_{3,2}^\theta(\xi\csc\theta,0)=2\theta H_2+\xi\pi H_1+\xi^2H_0\cot\theta-\frac{\xi^3\pi H(0)}{6}+\frac{\xi^4\tilde{H}_{-2}}{12}(3\cot\theta+\cot^3\theta)
+O(\xi^{\alpha_{min}+3}\ln\xi)+o(\xi^4)
\end{equation}
with omitted terms given in Eqs.~(\ref{e:3B},\ref{e:3C}).
As in the 2D case, we can now treat a general point (polar coordinates $(r,\theta)$) near an edge of angle $\omega$ (and hence solid
angle $2\omega$) as a sum of two such contributions, leading to
\begin{eqnarray}
M_{3,2}^\omega(r,\theta)&=&M_{3,2}^\theta(r,0)+M_{3,2}^{\theta'}(r,0)\nonumber\\
&=&2\omega H_2+r\pi H_1\left[\sin\theta+\sin\theta'\right]+r^2 H_0\left[\sin\theta\cos\theta+\sin\theta'\cos\theta'\right]-\frac{\pi r^3 H(0)}{6}[\sin^3\theta+\sin^3\theta']\\
&&+\frac{r^4\tilde{H}_{-2}}{12}[3\sin^3\theta\cos\theta+\sin\theta\cos^3\theta+3\sin^3\theta'\cos\theta'+\sin\theta'\cos^3\theta']\nonumber
+O(r^{\alpha_{min}+3}\ln r)+o(r^4)
\end{eqnarray}
where $\theta'=\omega-\theta$.
\end{widetext}

\begin{figure}
\centerline{\includegraphics[width=280pt]{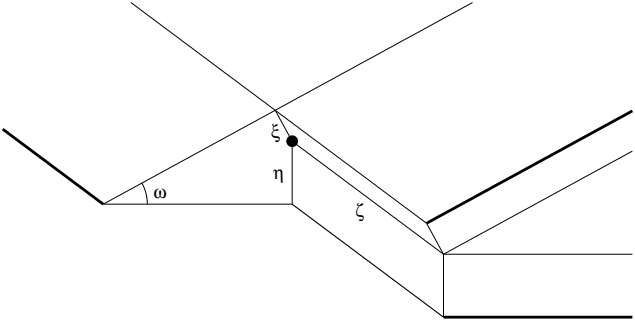}}
\caption{Geometry for the connectivity mass of a 3D corner, with the node at the indicated point and the section between it and the corner cut away.
The quantities $\xi$, $\eta$, $\zeta$ denote the perpendicular distances to the three faces. The domain is split into eight pieces, uniquely labelled by
which of these are relevant, for example $M_{\xi\eta\zeta}$ is the connectivity mass of the piece that was cut away, and $M_{\xi\eta}$ is a semi-infinite
kite shaped prism.\label{f:wedge3d}}
\end{figure}

Finally we consider a node near a right angled vertex, with angle and solid angle $\omega$, and located at $(r,\theta,\zeta)$ in cylindrical coordinates; the distance to
the angled planes are as before $\xi=r\sin\theta$ and $\eta=r\sin\theta'$: See Fig.~\ref{f:wedge3d}.  The connectivity mass is obtained by combining previous results 
for eight regions, for which we keep terms up to and including third order in the small quantities $\xi$, $\eta$ and $\zeta$:
\begin{equation}
M_{3,3}^\omega(r,\theta,\zeta)=M_.+M_\xi+M_\eta+M_\zeta+M_{\xi\eta}+M_{\xi\zeta}+M_{\eta\zeta}+M_{\xi\eta\zeta}
\end{equation}
Here,
\begin{equation}
M_.=\omega H_2
\end{equation}
is from the interior region obtained by translating the vertex so that it coincides with the node,
\begin{equation}
M_\xi=\frac{\pi}{2}\left[\xi H_1-\frac{\xi^3}{6}H(0)+\ldots\right]
\end{equation}
is from a quarter slab of width $\xi$, and similarly $M_\eta$; see the face contribution above.
\begin{equation} 
M_\zeta=\omega\left[\zeta H_1-\frac{\zeta^3}{6}H(0)+\ldots\right]
\end{equation}
is from a similar slab with an angle $\omega$ rather than $\pi/2$.
\begin{eqnarray}
M_{\xi\eta}&=&\frac{1}{2}\left[M_{C3}(\theta,\xi)+M_{C3}(\theta',\eta)\right]\nonumber\\
&=&\frac{H_0}{2}\left[\xi^2\cot\theta+\eta^2\cot\theta'\right]+\ldots
\end{eqnarray}
is from a semi-infinite strip with cross-section formed by two right angled triangles with common hypotenuse $r$ and angles $\theta$ and $\theta'$ respectively.
\begin{eqnarray}
M_{\xi\zeta}&=&\frac{1}{2}\left[M_{C3}(\arctan(\xi/\zeta),\xi)+M_{C3}(\arctan(\zeta/\xi),\zeta)\right]\nonumber\\
&=&H_0\xi\zeta+\ldots
\end{eqnarray}
is from a semi-infinite strip with rectangular cross-section of lengths $\xi$ and $\zeta$, that may be split into two right-angled triangles along the diagonal, and similarly
for $M_{\eta\zeta}$.  Finally
\begin{equation}
M_{\xi\eta\zeta}=\frac{H(0)}{2}\zeta r^2\left[\sin\theta\cos\theta+\sin\theta'\cos\theta'\right]+\ldots
\end{equation}
is from a prism with the same cross-section as $M_{\xi\eta}$ and length $\zeta$; since its extent is small in all directions its contribution (to third order)
is given by $H(0)$ multiplied by its volume.  Putting this together and expressing $\xi$ and $\eta$ in terms of $r$ and $\theta$ we have
\begin{widetext}
\begin{eqnarray}
M_{3,3}^\omega(r,\theta,\zeta)&=&\omega H_2+\left[\frac{\pi}{2}r(\sin\theta+\sin\theta')+\omega\zeta\right]H_1+\left[\frac{r^2}{2}(\sin\theta\cos\theta+\sin\theta'\cos\theta')+r\zeta(\sin\theta+\sin\theta')\right]H_0\nonumber\\
&&-\left[\pi r^3(\sin^3\theta+\sin^3\theta')-6r^2\zeta(\sin\theta\cos\theta+\sin\theta'\cos\theta')+2\omega\zeta^3\right]\frac{H(0)}{12}+o(r^3,\zeta^3)
\end{eqnarray}
\end{widetext}
where again $\theta'=\omega-\theta$.  As expected, we have $M_{3,2}^\omega(r,\theta)=2M_{3,3}^\omega(r,\theta,0)$ and $M_{3,1}(r)=M_{3,2}^\pi(r,\pi/2)$.

\section{Connection probability}\label{s:prob}
\subsection{Separation into boundary components}\label{s:sep}
Having obtained the connectivity mass in the vicinity of various boundaries in two and three dimensions,
we are now in a position to evaluate Eq.~(\ref{e:fc}) asymptotically (using Laplace's method)
for large $\rho$, and system size $L$, summing the dominant bulk and/or boundary contributions leading to Eq.~(\ref{e:SGF}).  We do not have a fully rigorous justification 
for this separation, however the neglect of contributions from intermediate regions may be justified as follows
(in two dimensions; we expect three dimensions to be similar):

Split the integration region $\cal V$ of the outer
integral appearing in Eq.~(\ref{e:fc}) into regions by lines
a distance $\epsilon_1$ and a distance $\epsilon_2$
from the boundary.  We will take $\rho$ and $L$ large, then choose $\frac{\ln\rho}{\rho}\ll\epsilon_1\ll 1$ and $1\ll\epsilon_2\ll L$.  Then, the contribution from the
intermediate regions (ie a distance from the boundary between $\epsilon_1$ and $\epsilon_2$) can be estimated, and is always of a lower order than at least one
of the main (corner, edge, bulk) contributions.  For example, the edge contribution in two dimensions (Eq.~(\ref{e:edge}) below) is of the form
\begin{equation}
P_{2,1}=Pe^{\rho\pi H_1}(1+o(1))
\end{equation}
where $P$ is the perimeter, proportional to $L$; this corresponds to a region in which the distance to an edge is less than $\epsilon_1$, but the distance to other edges is
greater than $\epsilon_2$.  Comparing with the bulk and corner contributions, this dominates when
\begin{equation}
\rho^{-1}e^{\rho(\pi-\omega)H_1}\lesssim L \lesssim\rho^{-1} e^{\rho\pi H_1}
\end{equation}   
There are three intermediate regions, where one or both of these distances is in $[\epsilon_1,\epsilon_2]$.  Taking the region where both are in this
range $I_{wb}$ near a corner of angle $\omega$ for example, we can estimate it by
\begin{equation}
I_{wb}\lesssim\rho e^{-\rho(\omega H_1+2\epsilon_1\sin(\omega/2) H_0)}
\end{equation}
using the connectivity mass at the point closest the corner (an angle $\omega/2$), Eq.~(\ref{e:wedge}).    So, we find that under our assumptions $P_{2,1}\gg I_{wb}$.
The other combinations of regimes may be estimated similarly, leading to the conclusion that in all cases, one of the bulk, edge or corner contributions dominates all three
intermediate contributions.  We expect a similar analysis to work in three dimensions also. So formally we conjecture that (compare with Eq.~\ref{e:SGF})
\begin{equation}\label{e:limit}
\frac{1-P_{fc}}{\sum_i\sum_{b\in{\cal B}_i}\rho^{1-i} G_{d,i}^{(b)} V_b e^{-\rho\Omega_b H_{d-1}}}\to 1
\end{equation}
in any limit where both $\rho$ and $L$ go to infinity.  Including terms in the denominator that are subleading in $\rho$ will not change the result, but should improve
the rate of convergence.  

\subsection{Polygons}\label{s:P2D}
We now present the results of Laplace's method for expanding Eq.~(\ref{e:fc}) for large $\rho$ in the two dimensional case; three dimensions is considered in the next section.
For convex polygons we have the following results from Sec.~\ref{s:wedge} above:
\begin{eqnarray}
M_{2,0}&=&2\pi H_1\nonumber\\
M_{2,1}(r)&=&\pi H_1+2rH_0\\
&&+\frac{r^3}{3}\left[\tilde{H}_{-2}+a_1\ln\left(\frac{2e^{5/6}}{|a_1|r}\right)\right]+o(r^3)\nonumber\\
M_{2,2}(\omega;r,\theta)&=&\omega H_1+r[\sin\theta+\sin\theta']H_0\nonumber\\
&&+\frac{H(0)r^2}{2}[\sin\theta\cos\theta+\sin\theta'\cos\theta']\nonumber\\
&&+O(r^3\ln r,r^{2+\alpha_{min}})
\nonumber
\end{eqnarray}
where $\omega$ is the angle of the corner, $(r,\theta)$ are polar coordinates of the node position, and other symbols and details are given in the above section.
The argument (as yet only semi-rigorous) is that for combined limits $\rho\to\infty$, $L\to\infty$ so that $P_{fc}\to 1$, a sum of boundary contributions takes
into account correctly the connectivity mass at locations of order $r_0$ from the boundary, which is not explicitly estimated above.  We have
\begin{equation}
P_{fc}=1-P_{2,0}-P_{2,1}-\sum_{corners} P_{2,2}
\end{equation}
where the corner contributions are separated out to allow for differing angles, while the bulk and edge contributions involve only the total area and perimeter
respectively.  The bulk contribution is
\begin{eqnarray}
P_{2,0}&=&\rho \int_{bulk} e^{-\rho M_{bulk}} dx dy\nonumber\\
&=&\rho A e^{-2\pi\rho H_1}
\end{eqnarray}
Here, $A$ is the area.  The edge contribution is (using $y$ to denote displacement along the edge and $x$ normal to it)
\begin{widetext}
\begin{eqnarray}\label{e:edge}
P_{2,1}&=&\rho \int_{edge} e^{-\rho M_{2,1}(x)} dx dy\nonumber\\
&=& \rho P \int_0^\infty e^{-\rho\left[[\pi H_1+2x H_0+\frac{x^3}{3}\left[\tilde{H}_{-2}+a_1\ln\left(\frac{2e^{5/6}}{|a_1|x}\right)\right]+O(x^5\ln x)\right]} dx\nonumber\\
&=& \rho P e^{-\rho\pi H_1} \int_0^\infty e^{- 2\rho H_0 x}\left[1-\frac{\rho x^3}{3}\left[\tilde{H}_{-2}+a_1\ln\left(\frac{2e^{5/6}}{|a_1|x}\right)\right]+O(x^5\ln x)\right]
dx\nonumber\\
&=&P e^{-\rho \pi H_1}\left[\frac{1}{2H_0}-\frac{\tilde{H}_{-2}+a_1\left[\gamma+\ln\left(\frac{4\rho H_0}{e|a_1|}\right)\right]}{8\rho^2H_0^4}+O(\rho^{-4}\ln\rho)\right]
\end{eqnarray}
Here, $P$ is the perimeter and $\gamma$ is Euler's constant.  Each corner contribution is
\begin{eqnarray}
P_{2,2}^\omega&=&\rho \int_{w} e^{-\rho M_{2,2}^\omega(r,\theta)} rdrd\theta\nonumber\\
&=&\rho \int_0^\omega d\theta\int_0^\infty rdr e^{-\rho\left[\omega H_1+rH_0(\sin\theta+\sin\theta')
+\frac{H(0) r^2}{2}(\sin\theta\cos\theta+\sin\theta'\cos\theta')+\frac{\tilde{H}_{-2}r^3}{6}(\sin^3\theta+\sin^3\theta')+O(r^{\alpha_{min}+2}\ln r,r^5)\right]}\nonumber\\
&=&\rho e^{-\rho\omega H_1} \int_0^\omega d\theta \int_0^\infty r dr e^{-\rho rH_0(\sin\theta+\sin\theta')}\nonumber\\
&&\left[1-\frac{\rho H(0) r^2}{2}(\sin\theta\cos\theta+\sin\theta'\cos\theta')-\frac{\rho\tilde{H}_{-2}r^3}{6}(\sin^3\theta+\sin^3\theta')+O(r^{\alpha_{min}+2}\ln r,r^4)\right]\label{e:P22}\\
&=& e^{-\rho\omega H_1} \int_0^\omega d\theta\nonumber\\
&&\left[\frac{1}{\rho H_0^2(\sin\theta+\sin\theta')^2}
-\frac{3H(0)(\sin\theta\cos\theta+\sin\theta'\cos\theta')}{\rho^2H_0^4(\sin\theta+\sin\theta')^4}-\frac{4\tilde{H}_{-2}(\sin^3\theta+\sin^3\theta')}{\rho^3H_0^5(\sin\theta+\sin\theta')^5}+O(\rho^{-\alpha_{min}-2}\ln\rho,\rho^{-4})\right]\nonumber\\
&=&e^{-\rho\omega H_1}\left[\frac{1}{\rho H_0^2\sin\omega}-\frac{H(0)(2\cos\omega+1)}{\rho^2H_0^4\sin^2\omega}-\frac{2\tilde{H}_{-2}}{\rho^3H_0^5\sin\omega}+O(\rho^{-\alpha_{min}-2}\ln\rho,\rho^{-4})\right]\nonumber
\end{eqnarray}

\subsection{Polyhedra}\label{s:P3D}
We can perform the same analysis on 3D shapes, using the results of Sec.~\ref{s:3D}.  We find
\begin{equation}
P_{fc}=1-P_{3,0}-P_{3,1}-\sum_{edges} P_{3,2}-\sum_{vertices} P_{3,3}
\end{equation}
where as above the edge and corner contributions are considered separately to allow for different angles, while the bulk and surface involve only
the total volume and surface area respectively.  The bulk, surface and edge contributions are, respectively,
\begin{eqnarray}
P_{3,0}&=&\rho V e^{-4\pi \rho H_2}\\
P_{3,1}&=&Se^{-2\pi \rho H_2}\left[\frac{1}{2\pi H_1}+\frac{2\pi H(0)}{\rho^2(2\pi H_1)^4}+\sum_\alpha\frac{a_\alpha \Gamma(2+\alpha)}{\rho^{2+\alpha}(2\pi H_1)^{4+\alpha}}
+O(\rho^{-5})\right]\\
P_{3,2}^\omega&=&Le^{-2\omega \rho H_2}\left[\frac{1}{\rho\pi^2H_1^2\sin\omega}-\frac{2H_0}{\rho^2\pi^4H_1^4}\frac{2\cos\omega+1}{\sin^2\omega}
+\frac{2\pi H(0)}{\rho^3\pi^5H_1^5\sin\omega}+O(\rho^{-\alpha_{min}-3}\ln\rho,\rho^{-4})\right]
\end{eqnarray}
where $V$ is the volume, $S$ the surface area and $L$ the length of an edge.  For a right-angled vertex of angle $\omega$ we have using the same approach
\begin{eqnarray}
P_{3,3}^\omega&=&\rho\int_0^\infty d\zeta\int_0^\omega d\theta\int_0^\infty rdr\\
&&e^{-\rho\left[\omega H_2+(\frac{\pi}{2}r(\sin\theta+\sin\theta')+\omega\zeta)H_1
+(\frac{r^2}{2}(\sin\theta\cos\theta+\sin\theta'\cos\theta')+r\zeta(\sin\theta+\sin\theta'))H_0+O(r^3,\zeta^3)\right]}\nonumber\\
&=&\rho\int_0^\infty d\zeta \left[\frac{e^{-\rho\omega(H_2+\zeta H_1)}}{\rho^2(\frac{\pi}{2}H_1+\zeta H_0)^2\sin\omega}
-\frac{H_0e^{-\rho\omega(H_2+\zeta H_1)}}{\rho^3(\frac{\pi}{2}H_1+\zeta H_0)^4}\frac{2\cos\omega+1}{\sin^2\omega}+O(\rho^{-4})\right]\nonumber
\end{eqnarray}
Noting again that $\rho$ is large and hence that only small $\zeta$ contribute, we expand the denominators in positive powers of $\zeta$ and integrate to give
\begin{equation}
P_{3,3}^\omega=e^{-\omega \rho H_2}\left[\frac{4}{\rho^2H_1^3\pi^2\omega\sin\omega}-\frac{16H_0}{\rho^3 H_1^5\pi^4}
\frac{\pi \sin\omega+2\omega\cos\omega+\omega}{\omega^2\sin^2\omega}+O(\rho^{-4})\right]
\end{equation}
\end{widetext}

\subsection{Leading and nonleading terms}\label{s:nonleading}
Comparing the 2D and 3D results of the previous sections with the geometrical factor
Eq.~(\ref{e:SGF}) we find the quantities given in Tab.~\ref{t:G}, which are remarkably
simple and general, and one of the main results of this paper.  In particular, the geometrical factor depends only on the connection
function via the $-i$ power of a single integral, namely $H_{d-2}$.

The nonleading terms involve smaller moments and the $a_\alpha$, that is, behaviour of the connection function near the origin.
Comparing the leading and second terms in the $P_{d,i}^\omega$ and noting that $H_m$ scales as $r_0^{m+1}$ in terms of a typical length scale
$r_0$,  we find they are the same order of magnitude if $\rho r_0^d$ is of order unity.  Physically this corresponds both to the average degree and
to the argument of the exponentials.  Thus for densities much above this, the terms in the expansions decrease rapidly, as we expect.

There are, however a few caveats.  The coefficients of the higher order terms may increase.  This is very common in asymptotic results; formally
the series may not converge, but in practice the first few terms remain a useful approximation of the function at values of the variable (density)
for which they decrease.

A more serious issue occurs for sharp corners.  The value of density at which the first two terms of the 2D corner contribution $P_{2,2}^\omega$,
Eq.~(\ref{e:P22}), are equal is
\begin{equation}
\rho_{eq}=\frac{H(0)}{H_0^2}\frac{2\cos\omega+1}{\sin\omega}
\end{equation}
which increases for sharp angles (small $\omega$), and also angles approaching $\pi$.    Thus care must be taken when
approximating $P_{2,2}^\omega$ at moderate densities.  The optimal angle is $2\pi/3$, for which the second term vanishes; this corresponds to
a hexagonal domain, popular in cellular networks.  The same holds for $P_{3,2}^\omega$; for $P_{3,3}^\omega$ the term vanishes at a
slightly higher angle of approximately $2.56125$ (for example, close to a hendecagonal prism).

\section{Curvature effects}\label{s:curv}
The previous calculations may be extended to geometries with curved boundaries.  If the boundary is smooth, then at sufficiently large system size $L$,
we can assume that the radius of curvature is much greater than the connection range, and so treat the curvature as a small parameter.  In two dimensions
a generic smooth boundary may be taken to have equation
\begin{equation}
z=\frac{\kappa x^2}{2}+O(x^3)
\end{equation}
and we place a node at $x=0$, $z=r>0$.  This convention makes the curvature $\kappa>0$ for convex domains.  Neglecting terms of order $\kappa^2$ and
$r^3$ for consistency, we find the curvature correction
\begin{eqnarray}
-M_{2,1}^\kappa(r)&=&\int_{-\infty}^\infty dx\int_0^{\kappa x^2/2}dz H(\sqrt{x^2+(z-r)^2})\nonumber\\
&=&\int_{-\infty}^\infty dx\int_0^{\kappa x^2/2}dz\left[H(x)+\frac{r^2}{2x}H'(x)+\ldots\right]\nonumber\\
&=&\kappa\left[H_2+\frac{r^2}{2}H'_1+\ldots\right]\\
&=&\kappa\left[H_2-\frac{r^2}{2}H_0+\ldots\right]\nonumber
\end{eqnarray}
using the integration by parts formula following Eq.~(\ref{e:Hm}).  Thus we update the calculation in Eq.~(\ref{e:edge}) to obtain
\begin{equation}
P_{2,1}=Pe^{-\rho(\pi H_1-\kappa H_2)}\left[\frac{1}{2H_0}-\frac{\kappa}{8\rho H_0^2}+O(\rho^{-2})\right]
\end{equation}
Notice that the curvature affects the exponential, hence reducing the effective angle slightly below $\pi$.  However the leading order geometrical factor remains unchanged.

In three dimensions, the corresponding leading order expression for the boundary is
\begin{equation}
z=\frac{1}{2}\left(\kappa_1x^2+\kappa_2y^2\right)+\ldots
\end{equation}
where $(\kappa_1,\kappa_2)$ are principal curvatures and $(x,y)$ displacements in the corresponding mutually orthogonal directions.  Using polar coordinates we have
\begin{eqnarray}
-M_{3,1}^\kappa(r)&=&\int_0^\infty \rho d\rho\int_0^{2\pi}d\theta\int_0^{(\kappa_1\cos^2\theta+\kappa_2\sin^2\theta)/2}dz \nonumber\\
&&\times H(\sqrt{\rho^2+(z-r)^2})\nonumber\\
&=&\int_0^\infty \rho d\rho\int_0^{2\pi}d\theta\int_0^{(\kappa_1\cos^2\theta+\kappa_2\sin^2\theta)/2}dz \nonumber\\
&&\times \left[H(\rho)+\frac{r^2}{2}H'(\rho)+\ldots\right]\nonumber\\
&=&\pi\kappa\left(H_3+\frac{r^2}{2}H'_2+\ldots\right)\\
&=&\pi\kappa(H_3-r^2H_1+\ldots)\nonumber
\end{eqnarray}
where $\kappa=(\kappa_1+\kappa_2)/2$ is the mean curvature.  From this we find
\begin{equation}
P_{3,1}=Se^{-\pi\rho(2H_2-\kappa H_3)}\left[\frac{1}{2\pi H_1}-\frac{\kappa}{\rho(2\pi H_1)^2}+O(\rho^{-2})\right]
\end{equation}
which has a similar structure to the two dimensional case.  Note also that it depends only on the mean curvature $\kappa$ and not the individual principal curvatures.

\section{Comparison with previous results and numerics}\label{s:ex}

The above expressions for $P_{fc}$ require only a few specific integrals of $H(r)$ for its evaluation, which for commonly used connection functions are given in Tab.~\ref{t:ex}.
Note that the expressions are valid whenever $\tilde{H}_{s-1}$ is defined; for the soft disk/annulus models, the contribution for negative $s$ comes from the discontinuity(-ies), while
in the Rayleigh fading case, from continuation of the integration by parts expressions. In the latter, $\tilde{H}_{-2}$ converges only for $\eta>1$ and $\tilde{H}_{-4}$ converges
only for $\eta>3$.  Some specific values for $\eta=2$ are given in Tab.~\ref{t:eta=2}.

We now compare the general results found above with geometric factors in specific cases studied previously, finding agreement with the above results.
The Rayleigh SISO model was considered in Ref.~\cite{CDG12a}, giving, for $\eta=2$,
\begin{equation}\label{e:gend}
G_{d,i}^{\pi/2}=2^{i(i-1)}\left(\frac{\beta}{\pi}\right)^{\frac{i(d-1)}{2}}
\end{equation}
for bulk, edges/faces and purely right angled corners in either two or three dimensions.    A general angle and path loss exponent was also considered in two
dimensions:
\begin{equation}
G_{2,2}^\omega=\frac{\beta^{2/\eta}}{\sin\omega \Gamma(1+\eta^{-1})^2}
\end{equation}
The earlier paper~\cite{CDG12b} gave special cases of these, namely $G_{3,i}^{\pi/2}$ and $G_{2,i}^\omega$, with a typo for $G_{2,1}^\omega$.  The other paper
with directly comparable results is Ref.~\cite{CGD14}.  Here, the model is $2\times 2$ MIMO with $\eta=2$, for which $H_1$ and $H_2$ are given in Tab.~\ref{t:eta=2}.
Finally, a circular or spherical boundary and Gaussian connection function were considered in Ref.~\cite{DGG}.  In all cases the results agree with
the more general expressions herein.

\begin{figure}[t]
\centerline{\includegraphics[width=250pt]{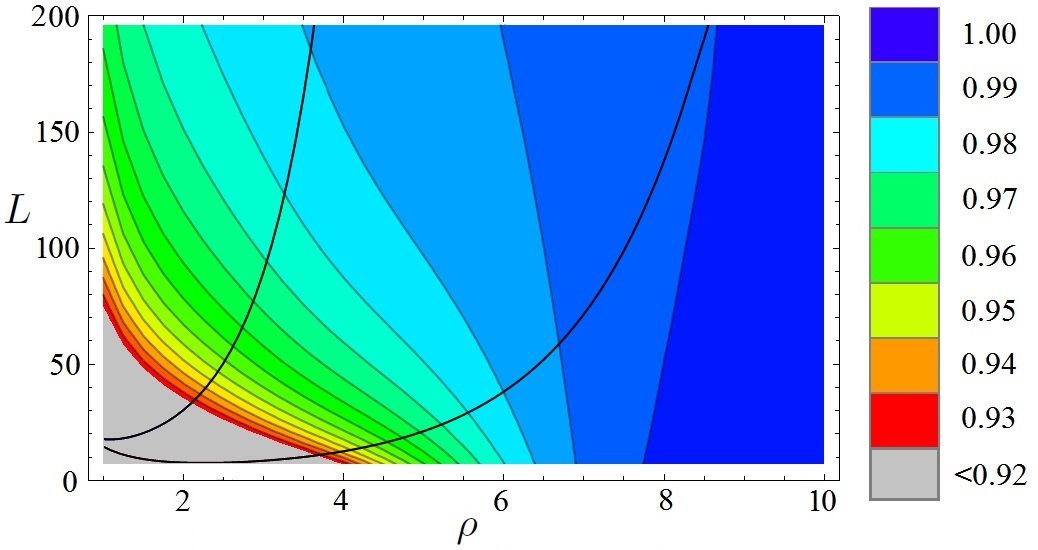}} 
\caption{Test of Eq.~(\protect\ref{e:limit}) for a square domain of side length $L$ and node density $\rho$.  The expression should tend to $1$ in any limit for which $\rho\to\infty$ and
$L\to\infty$.  The boundary between regions in which the corners (right), edges (middle) and bulk (left) dominate are shown in black; these appear to have no effect, thus illustrating the
uniformity of the expansion.\label{f:eq50}}
\end{figure}

\begin{figure}[t]
\centerline{\includegraphics[width=250pt]{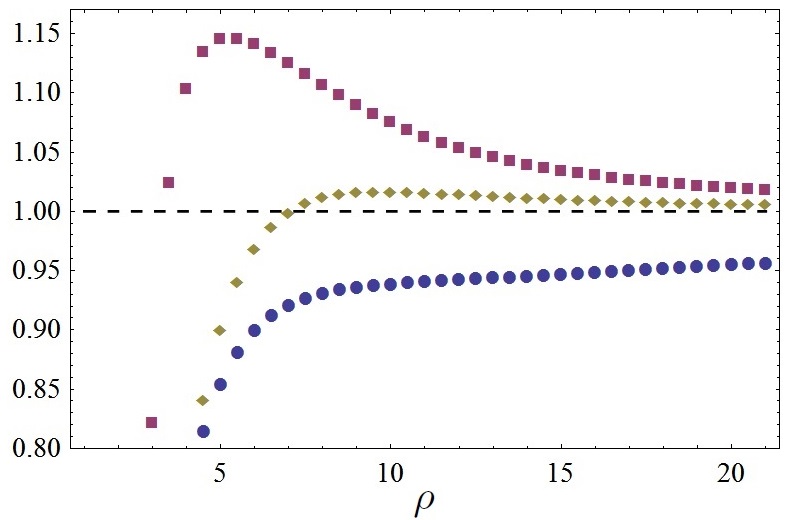}}
\caption{The ratio $\frac{1-P_{fc}-P_{2,0}-P_{2,1}}{P_{2,2}}$ for the square considered in Fig.~\protect\ref{f:eq50} and side length $L=10$, showing
convergence as more terms are included in $P_{2,2}$: Blue circles, purple squares, yellow diamonds to order $\rho^{-1}$, $\rho^{-2}$, $\rho^{-3}$ respectively.
\label{f:corner}}
\end{figure}

\begin{figure*}[t]
\centerline{\includegraphics[width=500pt]{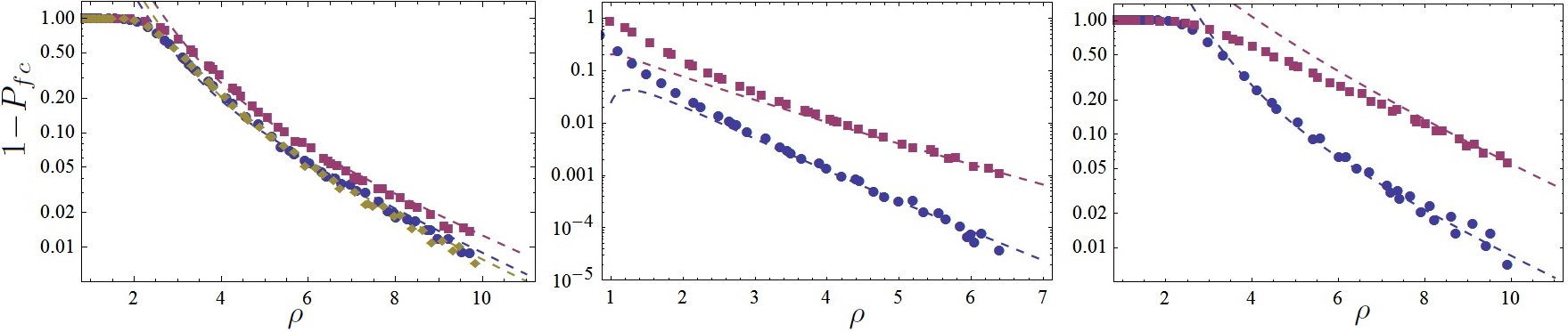}}
\caption{Full connection probability for a right triangle (side lengths 20 and 15) Left: Rayleigh with $\eta=3$ (top), $\eta=2$ (lower dark) and Rician with $\eta=2, K=4$ (lower light).
Centre: MIMO with $\eta=3$ (top) and $\eta=2$ (bottom).  Right: Soft annulus $r_-=1, r_+=2, a=1/3$ (top) and hard disk $r_0=1$ (bottom).\label{f:R2}}
\end{figure*}

We test Eq.~(\ref{e:limit}) for the case of a square domain of side length $L$ as shown in Fig.~\ref{f:eq50}.  The contributions from separate terms can be seen
in Fig.~\ref{f:corner}.  For both of these cases, the comparison is between the sum of boundary contributions and numerical integration of Eq.~(\ref{e:fc}).
A further test in Fig.~\ref{f:R2} compares the sum of boundary contributions with an ensemble of directly simulated random graphs for a variety of connection functions
for a triangular domain.  Thus it implicitly also confirms the validity of the assumptions undergirding  Eq.~(\ref{e:fc}) for these connection functions.

\section{Conclusion}\label{s:conc}
For random geometric graphs in finite geometries, the probability of (full) connection $P_{fc}$ can be conveniently approximated at high but finite node densities as a sum of
separable boundary contributions. Showing that these contributions can be obtained from a few moments for a very general class of connection functions and geometries, thus
vastly simplifying the evaluation of the relevant multidimensional integrals and hence the evaluation and design of ad-hoc wireless networks is the main contribution of the current
paper.  The results are in agreement with previous work and with numerics.

A number of previous works considered some examples where the above model and/or geometrical assumptions were relaxed, but not to the level of generality considered here:
\begin{itemize}
\item Dimensions other than $2$ or $3$: Eq.~(\ref{e:gend}) suggests further generalisation of the formulas and approach to $d>3$ might be possible (though perhaps
with fewer practical applications). On the other hand, for $d=1$ the connection probability is not dominated by that of an isolated node; it is quite likely in many
parameter regimes for the network to split into two or more large pieces.  For the unit disk model it is rather straightforward to calculate the probability of a gap of
given size, but for soft connection functions it remains open.

\item Anisotropic connections: These are of particular relevance in three dimensions (where antenna patterns are never exactly isotropic), and where beamforming is desirable to
mitigate interference from other nodes. The link pairwise connection
probability depends on orientation as well as mutual distance; see Refs.~\cite{CD13,GDC13b}.

\item Non-smooth boundaries:  Ref.~\cite{CGD14} considered a conical corner, which is not within the scope of the calculations presented in Sec.~\ref{s:curv}.
See also ``non-convex'' below.

\item Non-convex domains with a line-of-sight (LOS) condition: Examples have included keyhole geometries with~\cite{BDCR12} or without~\cite{GDC13c} reflections,
circular or spherical obstacles~\cite{DGG} and fractal domains~\cite{DGC14}.  In the latter, remarkably, it is found that $P_{fc}$ decreases toward zero in the limit
of high density. 
\end{itemize}
It would be interesting to extend the theory presented here to include these cases as well, providing a practical framework for understanding connectivity in diverse
spatial networks.

\section*{Acknowledgements} The authors would like to thank the directors of the Toshiba Telecommunications Research Laboratory for their support,
and Justin Coon, Ernesto Estrada and Martin Sieber for helpful discussions.

\bibliography{wireless}

\end{document}